\documentclass[pra,showpacs,twocolumn,aps,superscriptaddress,a4paper]{revtex4-1}
\usepackage{dcolumn,amssymb,amsmath,amsfonts,graphicx,latexsym,color}
\usepackage{epstopdf}

\usepackage{ulem}
\usepackage{epstopdf}
\usepackage{subfigure}
\usepackage{float}
\usepackage{mathrsfs}
\usepackage{bbold}
\usepackage{psfrag}
\usepackage{mathcomp}
\usepackage{verbatim}
\usepackage{multirow}
\usepackage{diagbox}
\usepackage[colorlinks,citecolor=blue]{hyperref}

\usepackage{simplewick}
\usepackage{CJK}
\begin{document}

%\begin{CJK*}{GBK}{song}

\title{Large-momentun tail of one-dimensional Fermi gases with spin-orbit coupling}

\author{Fang Qin}
\email{qinfang@ustc.edu.cn}
\affiliation{Shenzhen Institute for Quantum Science and Engineering and Department of Physics, Southern University of Science and Technology (SUSTech), Shenzhen 518055, China}
\affiliation{CAS Key Laboratory of Quantum Information, University of Science and Technology of China, Chinese Academy of Sciences, Hefei, Anhui 230026, China}

\author{Pengfei Zhang}
\affiliation{Walter Burke Institute for Theoretical Physics, California Institute of Technology, Pasadena, California 91125, USA}
\affiliation{Institute for Quantum Information and Matter, California Institute of Technology, Pasadena, California 91125, USA}

\author{Peng-Lu Zhao}
\affiliation{Shenzhen Institute for Quantum Science and Engineering and Department of Physics, Southern University of Science and Technology (SUSTech), Shenzhen 518055, China}

\date{\today}

\begin{abstract}
We study the contacts, large-momentum tail, radio-frequency spectroscopy, and some other universal relations for an ultracold one-dimensional (1D) two-component Fermi gas with spin-orbit coupling (SOC).
Different from previous studies, we find that the $q^{-8}$ tail in the spin-mixing (off-diagonal) terms of the momentum distribution matrix is dependent on the two SOC parameters in the laboratory frame for 1D systems, where $q$ is the relative momentum. This tail can be observed through time-of-flight measurement as a direct manifestation of the SOC effects on the many-body level.
Besides the traditional 1D even-wave scattering length, we find that two new physical quantities must be introduced due to the SOC.
Consequently, two new adiabatic energy relations with respect to the two SOC parameters are obtained. 
Furthermore, we derive the pressure relation and virial theorem at short distances for this system.
To find how the SOC modifies the large-momentum behavior, we take the SOC parameters as perturbations
since the strength of the SOC should be much smaller than the corresponding strength scale
of the interatomic interactions.
In addition, by using the operator product expansion method, we derive the asymptotic behavior of the large-momentum distribution matrix up to the $q^{-8}$ order and find that the diagonal terms of the distribution matrix include the contact of traditional 1D even-wave scattering length as the leading term and the SOC modified terms beyond the leading term, the off-diagonal term is beyond the subleading term and is corrected by the SOC parameters. 
We also find that the momentum distribution matrix shows spin-dependent and anisotropic features. 
Furthermore, we calculate the momentum distribution matrix in the laboratory frame for the experimental implication.
In addition, we calculate the high-frequency tail of the radio-frequency spectroscopy and find that the presence of the contact related to the center-of-mass momentum in the radio-frequency spectral is due to the SOC effects.
This paper paves the way for exploring the profound properties of many-body quantum systems with SOC in one dimension.
\end{abstract}

\maketitle

\section{Introduction}\label{1}

In the strong-coupling regime of cold atomic gases, a series of exact universal relations was established which set up a bridge between the microscopic short-distance correlations and the macroscopic thermodynamic properties of the many-body quantum system~\cite{Tan20081,Tan20082,Tan20083,Braaten20081,Braaten20082,Zhang2009,Platter2016}. These relations show that many thermodynamic properties are connected by a series of universal contact parameters which contain the information of the interaction effect in the large-momentum limit, and they are named contacts. 
These relations have already been successfully verified in experiments near the $s$-wave Feshbach resonance~\cite{Exp2010,Exp2012,Exp2013,Exp2019}.
Furthermore, the universal relations were also studied in other atomic systems such as the quantum gases with higher-partial-wave interactions~\cite{Yu2015,Zhou2016,Peng2016,Yoshida2015,Qin2016,Qin20182,Exp2016,Yu2017}, in one dimension~\cite{Barth2011,Cui20161,Cui20162,Patu2017,Yin2018} and two dimensions~\cite{Valiente2011,Valiente2012,Werner2012,Werner20122,Hofmann2D2012,Zhang2017,Peng2D2019,He1D2D2019}, with three-body correlations~\cite{Braaten3b2011,Braaten3b2014,Braaten3b2017,Yu20172,Yu20173,Nishida2013,Nishida2017}, near a Raman-dressed Feshbach resonance~\cite{Qin20181}, and with ultracold polar molecules~\cite{He2020}. 
Besides, the contact tensor was predicted in the axisymmetry-broken $p$-wave Fermi gases~\cite{tensor2016}. At the same time, Zhang $et~al.$ predicted the contact matrix and studied the direct connection between the contact matrix and the order parameter of a superfluid~\cite{Matrix2017}. 
The nuclear neutron-proton contact was introduced in nuclear physics~\cite{Weiss20151}, and the general nuclear contact matrices were defined~\cite{Weiss20152} in the context of generalized contacts for realistic interactions~\cite{Weiss2016,Weiss20171,Weiss20172,Weiss2018,Weiss2019}. There are also other works using Coulomb~\cite{Hofmann2013} and realistic atom-atom interactions~\cite{Valiente2020} with the generalized contact formalism.

The successful experimental realization of the synthetic coupling between atomic (pseudo) spin and momentum is another important progress in cold atomic gases~\cite{ExpSOC2011,ExpSOC20121,ExpSOC20122,ExpSOC20123,ExpSOC2013,ExpSOC2016,ExpSOC2018}.
This synthetic coupling is so-called spin-orbit coupling (SOC). 
In one dimension, there is only one type of SOC: The spin is only coupled to the motion of atoms along one spatial direction which is induced by two contour-propagating Raman beams~\cite{Goldman2014,Zhang2014,Zhai2012,Zhai2015}. 
This type of SOC has been  realized in experiments for both Bose and Fermi gases~\cite{ExpSOC2011,ExpSOC20121,ExpSOC20122,ExpSOC20123,ExpSOC2013}.
Importantly, the SOC can strongly affect the many-body properties~\cite{Xu2015,Yi2015}. 

Recently, Peng $et~al.$ studied the universal relations for the Fermi gases with a three-dimensional isotropic
SOC~\cite{PengSOC2018}. At the same time, the universal relations for the three-dimensional ultracold gases with an arbitrary type of SOC have been investigated~\cite{JieSOC2018,ZhangSOC2018}. Furthermore, the universal relations for spin-orbit-coupled Fermi gases in two dimensions have been derived \cite{PengSOC2019}. 
However, the one-dimensional (1D) Fermi gases with SOC has not been studied. 
Here, we investigate the 1D Fermi gases with SOC.
Different from the previous studies, we derive the asymptotic behavior of the large-momentum distribution matrix up to the $q^{-8}$ order ($q$ is the relative momentum) and find that the $q^{-8}$ tail in the spin-mixing (off-diagonal) terms of the momentum distribution matrix is dependent on the SOC parameters in the laboratory frame for 1D systems. This tail can be observed through time-of-flight measurement and it is a direct manifestation of the SOC effects on the many-body level.

In this paper, we focus on a many-body quantum system that exhibits
universal properties, i.e., the 1D two-component Fermi gas with SOC near a broad $s$-wave Feshbach resonance. 
Such a 1D system can be realized by applying tight transverse confinements to three-dimensional gases near $s$-wave Feshbach resonances, such as in the fermions of $^6$Li and $^{40}$K~\cite{Exp1D2004,Exp1D2008,Exp1D20101,Exp1D20102,Olshanii1998,Olshanii2003,Cui1D2012,Qin1D2017,Sala1D2012}.
First of all, we give the definition of the traditional even-wave contact for this system.
After that, we derive the universal relations for the spin-orbit-coupled Fermi gases in one dimension, including the adiabatic relations, pressure relation, and viral theorem. 
Besides the traditional even-wave scattering length, we find that two new physical quantities need to be introduced due to the SOC.
Using the operator product expansion (OPE) method, we derive the large-momentum tail of the
momentum distribution matrix. 
The momentum distribution matrix shows spin-dependent and anisotropic features. 
We find that the diagonal elements of the distribution matrix include the contact of the traditional even-wave scattering length as the leading term and the SOC modified terms beyond the leading term, the off-diagonal term is beyond the subleading term and is corrected by the SOC parameters. 
Furthermore, in order to discuss the experimental implications, 
we calculate the momentum distribution in the laboratory frame and find that the spin-mixing (off-diagonal) elements of the momentum distribution matrix at large momentum is directly modified by the SOC parameters and the tails which beyond the leading-order term have been changed, the effects 
can be captured by a time-of-flight measurement in experiments. 
In addition, we derive the high-frequency radio-frequency spectroscopy and find that the presence of the contact related to the center-of-mass momentum in the radio-frequency spectral is due to the SOC effects.

The paper is organized as the following: In Sec.~\ref{2}, we give the model Hamiltonian. 
In Sec.~\ref{3}, we calculate the two-body physics. In Sec.~\ref{4}, we give the definition of the traditional even-wave contact. Furthermore, we study some of the universal relations in Sec.~\ref{5}. Moreover, we calculate the large-momentum distribution tail in Sec.~\ref{6}. In addition, we calculate the high-frequency tail of the radio-frequency spectroscopy in Sec.~\ref{7}. Finally, we summarize in Sec.~\ref{8}.

\section{Model and noninteracting properties}\label{2}

With the pseudopotential approximation, the effective 1D Hamiltonian with SOC is given by ($\hbar=1$ throughout the paper)~\cite{Zhai2012,Zhai2015,Olshanii1998,Olshanii2003,Cui1D2012,Qin1D2017}
\begin{align}\label{eq:Hamiltonian-r}
H &= \sum_{\sigma=\uparrow,\downarrow} \int dx~\psi^{\dag}_{\sigma}(x)\left(- \frac{1}{2m}\frac{\partial^{2}}{\partial x^{2}} \right)\psi_{\sigma}(x) \nonumber\\
&~~+ \int dx~\Omega[e^{i2k_{0}x}\psi^{\dag}_{\uparrow}(x)\psi_{\downarrow}(x) + e^{-i2k_{0}x}\psi^{\dag}_{\downarrow}(x)\psi_{\uparrow}(x)] \nonumber\\
&~~+ g_{1D}^{}\int dx~\psi^{\dag}_{\uparrow}(x)\psi^{\dag}_{\downarrow}(x)\psi_{\downarrow}(x)\psi_{\uparrow}(x),
\end{align}
where $\psi_{\sigma}(x)$ is the field operator for the fermionic atoms, $x$ is the longitudinal atomic separation,  the Fermi atoms in state $|\uparrow\rangle$ are coupled to state $|\downarrow\rangle$ by the Raman laser with the strength $\Omega=\Omega_{R}/2$, $\Omega_{R}$ is the Rabi frequency, $2k_{0}$ is the momentum transfer during the two-photon processes, and $g_{1D}^{}$ is the even-wave coupling constant in 1D.

To remove the phase factor $e^{\pm i2k_{0}x}$ in the second term of Eq.~(\ref{eq:Hamiltonian-r}), we introduce two new atomic fields: $\psi_{\uparrow}(x)=\psi_{\uparrow}(x)e^{ik_{0}x}$ and $\psi_{\downarrow}(x)=\psi_{\downarrow}(x)e^{-ik_{0}x}$. 
Then, we can write the single-particle part of the Hamiltonian in the momentum space: $H_{0}=\sum_{k}\Psi_{k}^{\dag}{\cal H}_{0}\Psi_{k}$ with $\Psi_{k}=(a_{k,\uparrow},a_{k,\downarrow})^{T}$ and 
\begin{align}
{\cal H}_{0} = 
\left(\begin{array}{cc}
\frac{(k + k_{0})^{2}}{2m}  & \Omega \\
\Omega   &  \frac{(k - k_{0})^{2}}{2m}
\end{array}\right),
\end{align} where $a_{k,\sigma}$ is the field operator for the fermionic atoms in the momentum space.

Therefore, the inverse of the single-particle propagator matrix is given by~\cite{ZhangSOC2018,He2018,Stoof2009}
\begin{align}
&G^{-1}(q_{0},k) =  \nonumber\\
&-i\left(\begin{array}{cc}
q_{0} + i0^{+} - \frac{(k + k_{0})^{2}}{2m}  & -\Omega \\
-\Omega   &  q_{0} + i0^{+} - \frac{(k - k_{0})^{2}}{2m}
\end{array}\right),
\end{align} where $q_{0}$ is the total energy.
Furthermore, we have 
\begin{align}\label{eq:G0}
G(q_{0},k) 
&= \int_{0}^{\infty} dt~e^{iq_{0}t}\langle{\cal T}\Psi_{k}(t)\Psi_{k}^{\dagger}(0)\rangle \nonumber\\
&= \left(\begin{array}{cc}
G_{\uparrow\uparrow}(q_{0},k)  & G_{\uparrow\downarrow}(q_{0},k) \\
G_{\downarrow\uparrow}(q_{0},k)  &  G_{\downarrow\downarrow}(q_{0},k)
\end{array}\right),
\end{align} where ${\cal T}$ is the time-ordered operator and the elements are 
\begin{align}
&G_{\uparrow\uparrow}(q_{0},k) = \frac{i}{ q_{0} - \frac{(k + k_{0})^{2}}{2m} + i0^{+} - \frac{\Omega^{2}}{ q_{0} - \frac{(k - k_{0})^{2}}{2m} + i0^{+} }}, \\
&G_{\uparrow\downarrow}(q_{0},k) = G_{\downarrow\uparrow}(q_{0},k) = \nonumber\\
&~~\frac{i\Omega}{\left[ q_{0} - \frac{(k - k_{0})^{2}}{2m} + i0^{+} \right]\left[ q_{0} - \frac{(k + k_{0})^{2}}{2m} + i0^{+} \right] - \Omega^{2}}, \\
&G_{\downarrow\downarrow}(q_{0},k) = \frac{i}{ q_{0} - \frac{(k - k_{0})^{2}}{2m} + i0^{+} - \frac{\Omega^{2}}{q_{0} - \frac{(k + k_{0})^{2}}{2m} + i0^{+} }}.
\end{align}

To calculate the Feynman diagrams for simplicity, the interacting part of the Hamiltonian can be written as 
\begin{eqnarray}
H_{\text{int}}  
&=& g_{1D}^{}\int dx\left[\frac{1}{2}\Psi^{\dagger}(x)\epsilon^{\dagger}\Psi^{\dagger T}(x)\right]\left[\frac{1}{2}\Psi^{T}(x)\epsilon\Psi(x)\right],\nonumber\\
\end{eqnarray} where $\epsilon=-i\sigma_{y}$ is the two-by-two antisymmetric matrix, $\Psi(x)=[\psi_{\uparrow}(x),\psi_{\downarrow}(x)]^{T}$, and we have
\begin{align}
\frac{1}{2}\Psi^{T}(x)\epsilon\Psi(x) &= \frac{1}{2}[\psi_{\downarrow}(x)\psi_{\uparrow}(x) - \psi_{\uparrow}(x)\psi_{\downarrow}(x)] \nonumber\\
&= \psi_{\downarrow}(x)\psi_{\uparrow}(x).
\end{align}

We can also write the interaction part of the Hamiltonian in the momentum space~\cite{ZhangSOC2018,Yao2015,Nandkishore2013,Moon2013}: 
\begin{align}\label{eq:Hamiltonian-int}
H_{\text{int}} = g_{1D}^{}\sum_{k,k'}\left(\frac{1}{2}\Psi_{k,k'}^{\dagger}\epsilon^{\dagger}\Psi_{k,k'}^{\dagger T}\right)\left(\frac{1}{2}\Psi_{k,k'}^{T}\epsilon\Psi_{k,k'}\right),
\end{align} where $\Psi_{k,k'}=(a_{k,\uparrow}^{},a_{k',\downarrow}^{})^{T}$ and we have
\begin{align}
\frac{1}{2}\Psi_{k,k'}^{T}\epsilon\Psi_{k,k'} = \frac{1}{2}(a_{k',\downarrow}^{}a_{k,\uparrow}^{} - a_{k,\uparrow}^{}a_{k',\downarrow}^{}) = a_{k',\downarrow}^{}a_{k,\uparrow}^{}.
\end{align}

\section{Two-body physics}\label{3}

%%%%%%%%%%%%%%%%%%%%%%%%%%%%%%%%%%%%%%%%%%%%%%%%%%
\begin{figure}
\includegraphics[width=8cm]{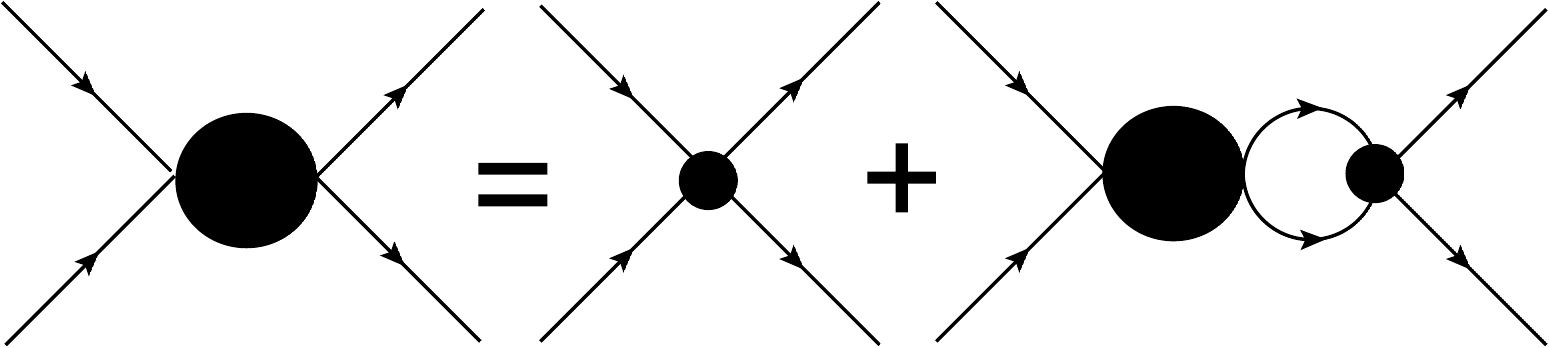}
\caption{Feynman diagrams for the $T$ matrix near a wide $s$-wave Feshbach resonance. The single line denotes the bare atom propagator matrix $G$. The black disk represents the $T$ matrix: $-iT_{s}$.
The black dot represents the interaction vertex: $-ig_{1D}^{}$.} \label{fig:Ts}
\end{figure}
%%%%%%%%%%%%%%%%%%%%%%%%%%%%%%%%%%%%%%%%%%%%%%%%%%

The two-body scattering process can be described by the two-body scattering amplitude or the scattering phase shift which can be determined by the two-body $T$ matrix in physics. 
In order to show the bare coupling constant $g_{1D}^{}$ for the even-wave case needs to be renormalized or not, it is necessary to calculate the two-body $T$ matrix at first.
Because the $T$ matrix is a physical quantity describing
effective interaction in the low-energy space irrelevant to
short-range physics.

We consider finite total momentum $Q$ for each two-body pairing state, so that an incoming state can be set as $|I_s\rangle=|Q/2 + k,\uparrow;Q/2 - k,\downarrow\rangle$ with two fermions of different species having momenta $Q/2 + k$ and $Q/2 - k$ to an outgoing state $|O_s\rangle=|Q/2 + k',\uparrow; Q/2 - k',\downarrow\rangle$ with two fermions having momenta $Q/2 + k'$ and $Q/2 - k'$.

As shown in Fig.~\ref{fig:Ts}, the two-body $T$ matrix is given by~\cite{Gurarie2007}
\begin{eqnarray}\label{eq:Ts}
-iT_{s}(q_{0},Q) = \frac{-ig_{1D}^{}}{ 1 - (-ig_{1D}^{})\Pi_{s}(q_{0},Q)},
\end{eqnarray}
where the polarization bubble is given by Refs.~\cite{ZhangSOC2018,Yao2015,Nandkishore2013,Moon2013} (the derivations are given in the Appendix),
\begin{align}\label{eq:bubble}
&~~\Pi_{s}(q_{0},Q) \nonumber\\
&= \int\frac{dpdp_{0}}{(2\pi)^2} \frac{1}{2} \text{Tr}\left[ G^{T}(p_{0},Q/2 + p)\epsilon G(q_{0}-p_{0},Q/2 - p)\epsilon^{\dagger} \right] \nonumber\\
&= \int\frac{dpdp_{0}}{(2\pi)^2} \frac{1}{2} \left[ G_{\uparrow\uparrow}(p_{0}, Q/2 + p)G_{\downarrow\downarrow}(q_{0}-p_{0},Q/2 - p) \right. \nonumber\\
&~~\left. + G_{\downarrow\downarrow}(p_{0},Q/2 + p)G_{\uparrow\uparrow}(q_{0}-p_{0},Q/2 - p) \right. \nonumber\\
&~~\left. - G_{\uparrow\downarrow}(p_{0},Q/2 + p)G_{\downarrow\uparrow}(q_{0}-p_{0},Q/2 - p) \right. \nonumber\\
&~~\left. - G_{\downarrow\uparrow}(p_{0},Q/2 + p)G_{\uparrow\downarrow}(q_{0}-p_{0},Q/2 - p) \right],
\end{align} $q_{0}^{}=Q^{2}/(4m) + k^{2}/m$, and $\text{Tr}$ denotes the trace over the spin degrees of freedom.

In the absence of the Raman coupling, the even-wave coupling constant is given by (the derivations are given in the Appendix)
\begin{eqnarray}\label{eq:g1D}
g_{1D}^{}  = - \frac{1}{m_{r}a_{1D}^{}},
\end{eqnarray} where $m_{r}=m/2$ is the reduced mass, the even-wave scattering length in 1D is given by~\cite{Olshanii1998}  
\begin{align}\label{eq:a1D}
a_{1D}^{} = - \frac{\ell_{\perp}^{2}}{2a_{s}} + \frac{{\cal C}\ell_{\perp}}{2}, 
\end{align} $a_{s}$ is the three-dimensional scattering length, ${\cal C}=1.4603$, $\ell_{\perp}=\sqrt{2/(m\omega_{\perp})}$, and $\omega_{\perp}$ is the transverse trapping frequency.

Note that Eq.~(\ref{eq:a1D}) can only be valid for either a broad Feshbach resonance or the field-free case. Furthermore, the confinement induced resonance should also be broad. Since the external confinement
is never perfectly harmonic nor isotropic, the possibility of inelastic resonances is possible. 
For example, a splitting of confinement-induced resonances has been observed in an anisotropic quasi-1D gas of Cs atoms in experiment~\cite{Exp1D20102}. Later, a theoretical model of the inelastic confinement-induced resonance was presented to describe the experimental observations~\cite{Sala1D2012}.

In the presence of the Raman fields, the polarization bubble with zero total momentum ($Q=0$) can be calculated as
\begin{align}\label{eq:bubbleSOC}
&\Pi_{s}(E_{0},0) = \frac{m^{3}\Omega^{2}}{2(mE_{0}k_{0}^{2}-k_{0}^{4}+m^{2}\Omega^{2})\sqrt{mE_{0}+i0^{+}-k_{0}^{2}}}  \nonumber\\
&~~+ \frac{mk_{0}^{2}}{4} \left[ \frac{\sqrt{mE_{0}+k_{0}^{2}+2\sqrt{mE_{0}k_{0}^{2}+m^{2}\Omega^{2}} } }{m^{2}\Omega^{2} + k_{0}^{2}(mE_{0} + \sqrt{mE_{0}k_{0}^{2}+m^{2}\Omega^{2}})} \right. \nonumber\\
&~~\left.+ \frac{\sqrt{mE_{0}+i0^{+}+k_{0}^{2}-2\sqrt{mE_{0}k_{0}^{2}+m^{2}\Omega^{2}} } }{m^{2}\Omega^{2} + k_{0}^{2}(mE_{0} - \sqrt{mE_{0}k_{0}^{2}+m^{2}\Omega^{2}})} \right],
\end{align} where $E_{0}=k^{2}/m$ and the $+i0^+$ terms are needed when $E_{0}<k_{0}^{2}/m$ and $mE_{0}+k_{0}^{2}<2\sqrt{mE_{0}k_{0}^{2}+m^{2}\Omega^{2}}$ such as $\sqrt{mE_{0}+i0^{+}-k_{0}^{2}}=i\sqrt{k_{0}^{2}-mE_{0}}$ and $\sqrt{mE_{0}+i0^{+}+k_{0}^{2}-2\sqrt{mE_{0}k_{0}^{2}+m^{2}\Omega^{2}} }=i\sqrt{2\sqrt{mE_{0}k_{0}^{2}+m^{2}\Omega^{2}} - (mE_{0}+k_{0}^{2})}$.

Note that the bubble (\ref{eq:bubbleSOC}) does not diverge and we do not need to have renormalization relation for $g_{1D}^{}$.

The even-wave scattering amplitude in 1D can be written as~\cite{Barth2011,Olshanii1998} $f_{1D}(k) = -1/(1+i\cot\delta_{k})$, where $\delta_{k}$ is the scattering phase shift.
With $T_{s}(k)=ikf_{1D}(k)/m_{r}$ and Eq.~(\ref{eq:bubbleSOC}), one can get
\begin{align}\label{eq:g1DSOC}
\cot\delta_{k}  &= \frac{k}{m_{r}}\left\{ - \frac{1}{g_{1D}^{}} - i\left[ \Pi_{s}(E_{0},0) - \frac{m_{r}}{k} \right] \right\}.
\end{align}

\section{Even-wave contact}\label{4}

%%%%%%%%%%%%%%%%%%%%%%%%%%%%%%%%%%%%%%%%%%%%%%%%%%
\begin{figure}
\includegraphics[width=8cm]{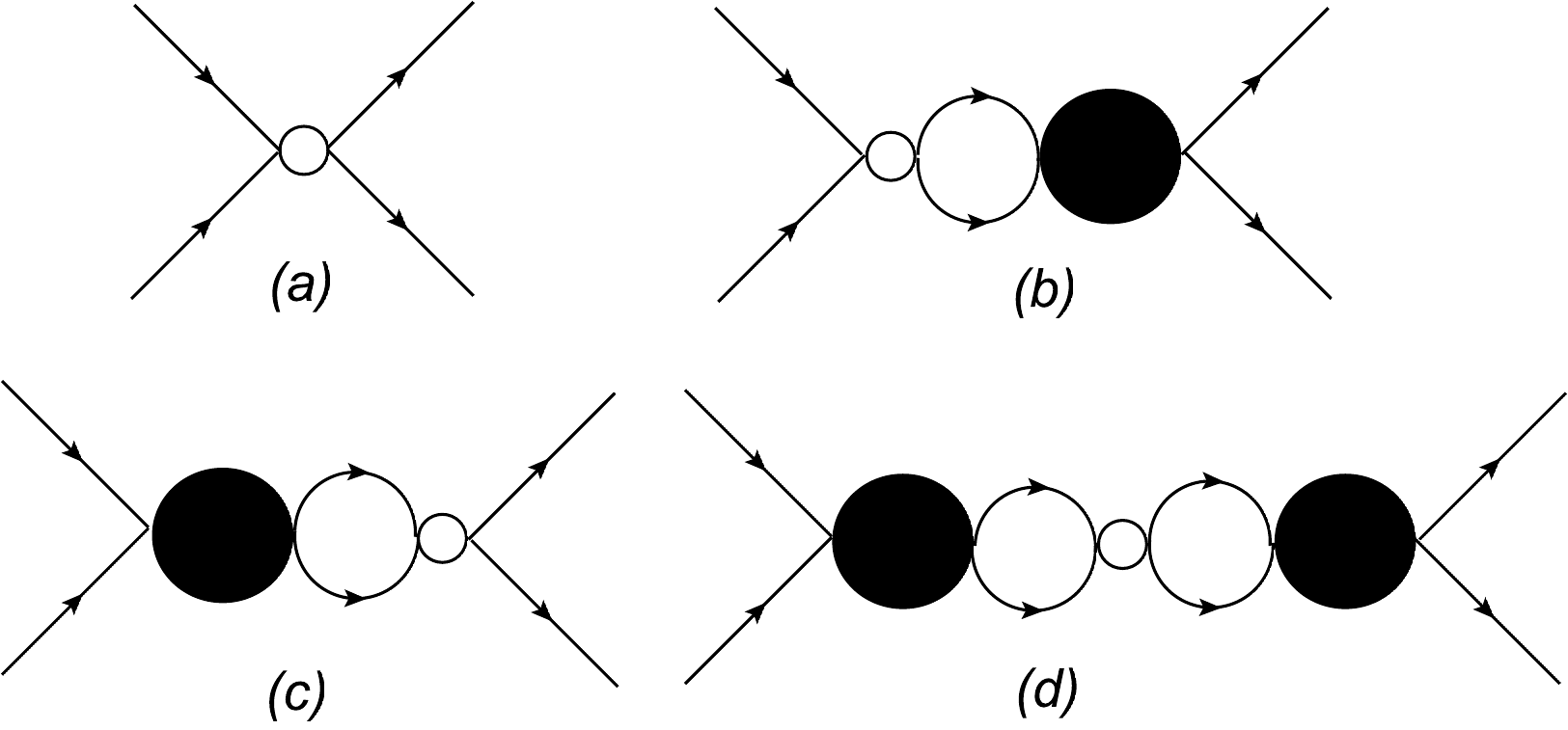}
\caption{Feynman diagrams for the matrix elements of the
two-atom local operator $\psi^{\dagger}_{\uparrow}(R) \psi^{\dagger}_{\downarrow}(R) \psi_{\downarrow}(R) \psi_{\uparrow}(R)$ and its derivatives $\psi^{\dagger}_{\uparrow}(R) \psi^{\dagger}_{\downarrow}(R) \left[i\partial_{t}+\partial_{R}^{2}/(4m) \right]^{j} \psi_{\downarrow}(R) \psi_{\uparrow}(R)$, $\psi^{\dagger}_{\uparrow}(R) \psi^{\dagger}_{\downarrow}(R) (-i\partial_{R})^{n} \psi_{\downarrow}(R) \psi_{\uparrow}(R)$, and $\psi^{\dagger}_{\uparrow}(R) \psi^{\dagger}_{\downarrow}(R) \left[i\partial_{t}+\partial_{R}^{2}/(4m) \right]^{j}(-i\partial_{R})^{n} \psi_{\downarrow}(R) \psi_{\uparrow}(R)$ with $j,n=1,2,3,\cdot\cdot\cdot$. The open dots represent the operators. \label{fig:TwoBodyOperatorRs}}
\end{figure}
%%%%%%%%%%%%%%%%%%%%%%%%%%%%%%%%%%%%%%%%%%%%%%%%%%

To define the 1D even-wave contact, we need to give the adiabatic relation~\cite{Barth2011}, 
\begin{align}
\frac{C_{a}}{2m} \equiv \frac{\partial E}{\partial a_{1D}^{}} = \int dR \left\langle \frac{\partial {\cal H}}{\partial a_{1D}^{}} \right\rangle , \label{eq:caE}
\end{align} where $E$ is the total energy of the many-body system, ${\cal H}$ is the density of the Hamiltonian (\ref{eq:Hamiltonian-r}), $C_{a}$ is the 1D even-wave contact, and we have used the following relations:
\begin{align}
\left\langle \frac{\partial {\cal H}}{\partial a_{1D}^{}} \right\rangle
&= \left\langle \frac{\partial {\cal H}}{\partial g_{1D}^{}} \right\rangle \frac{\partial g_{1D}^{}}{\partial a_{1D}^{}} \nonumber\\
&= \frac{m}{2}g_{1D}^{2} \langle \psi^{\dag}_{\uparrow}(R) \psi^{\dag}_{\downarrow}(R) \psi_{\downarrow}(R) \psi_{\uparrow}(R) \rangle. \label{eq:ca0}
\end{align}

Substituting Eq.~(\ref{eq:ca0}) into (\ref{eq:caE}), we get the expression for the contact $C_{a}$ as 
\begin{align}
C_{a} = m^{2} g_{1D}^{2} \int dR \langle \psi^{\dag}_{\uparrow}(R) \psi^{\dag}_{\downarrow}(R) \psi_{\downarrow}(R) \psi_{\uparrow}(R) \rangle, \label{eq:ca1}
\end{align}
where it is indicated that $C_{a}$ refers to the two-atom operator.

Furthermore, we calculate the expectation values of the
two-atom operator $\langle O_s| \psi^{\dag}_{\uparrow}(R) \psi^{\dag}_{\downarrow}(R) \psi_{\downarrow}(R) \psi_{\uparrow}(R) |I_s \rangle$ as shown in Fig.~\ref{fig:TwoBodyOperatorRs}~\cite{Qin20182}, 
\begin{eqnarray}
&& \langle O_s| \psi^{\dag}_{\uparrow}(R) \psi^{\dag}_{\downarrow}(R) \psi_{\downarrow}(R) \psi_{\uparrow}(R) |I_s \rangle \nonumber \\
&=& \sum_{j=a,b,c,d} \langle O_s| \psi^{\dag}_{\uparrow}(R) \psi^{\dag}_{\downarrow}(R) \psi_{\downarrow}(R) \psi_{\uparrow}(R) |I_s \rangle_{j} \nonumber\\
&=& [ 1 -iT_{s}(q_{0},Q) \Pi_{s}(q_{0},Q) ]^2. \label{eq:TwoBodyOperatorsR1}
\end{eqnarray}

Substituting Eq.~(\ref{eq:Ts}) into (\ref{eq:TwoBodyOperatorsR1}), we have
\begin{eqnarray}\label{eq:TwoBodyOperatorsR2}
\langle O_s| \psi^{\dag}_{\uparrow}(R) \psi^{\dag}_{\downarrow}(R) \psi_{\downarrow}(R) \psi_{\uparrow}(R) |I_s \rangle = \frac{[T_{s}(q_{0},Q)]^2}{g_{1D}^{2}}.
\end{eqnarray}

Substituting Eq.~(\ref{eq:TwoBodyOperatorsR2}) into (\ref{eq:ca1}), we get the even-wave contact as
\begin{align}
C_{a} = m^{2} \int dR~[T_{s}(q_{0},Q)]^{2}. \label{eq:ca}
\end{align}

\section{Universal relations}\label{5}

\subsection{Adiabatic relations}\label{5.1}

It is known that a contact can be defined to characterize the variation of energy as shown in Eq.~(\ref{eq:caE}).
When SOC is present, there are two new parameters $k_{0}$ and $\Omega$. 
Similar to how we deal with the scattering length $a_{1D}^{}$, one can define two new physical quantities $C_{\lambda}$ and $C_{\Omega}$ as
\begin{align}
C_{\lambda} &\equiv \int dR \left\langle \frac{\partial {\cal H}}{\partial k_{0}} \right\rangle,\label{eq:ck0} \\
C_{\Omega} &\equiv \int dR \left\langle \frac{\partial {\cal H}}{\partial \Omega} \right\rangle.\label{eq:cOmega}
\end{align}
Here, $C_{\lambda}$ and $C_{\Omega}$ refer to only single-atom operators which give nonzero matrix elements
in the single-atom sector. The momentum distribution under
single-particle states is just a $\delta$ function so that $C_{\lambda}$ and $C_{\Omega}$ will not
contribute to the large-momentum tail, which is different from $C_{a}$~\cite{ZhangSOC2018}. 
However, both $k_{0}$ and $\Omega$ have nonzero energy scales so that they would appear
in the pressure relation and viral theorem. 

Note that the idea of the contact is that it describes coalescence of particles.
$C_{\lambda}$ and $C_{\Omega}$ refer to single-particle properties, i.e., they are
obviously nonzero even for a single-particle or a noninteracting
Fermi gas. Therefore, one cannot call them contacts, and we call them new physical quantities here.

\subsection{Pressure relation}\label{5.2}

For a uniform gas, the pressure relation can be derived following the expression of the Helmholtz free energy density ${\cal F}=F/L$ which can be expressed in terms of~\cite{Braaten20082,Barth2011,Cui20161}
\begin{align}
&~~{\cal F}(T, n_{\uparrow}, n_{\downarrow}, a_{1D}^{}, k_{0}, \Omega)  \nonumber\\
&= \frac{k_{F}^{3}}{2m} f\left(\frac{2mT}{k_{F}^{2}}, \frac{n_{\uparrow}}{k_{F}}, \frac{n_{\downarrow}}{k_{F}}, a_{1D}^{} k_{F}, \frac{k_{0}}{k_{F}}, \frac{2m\Omega}{k_{F}^{2}} \right), \label{eq:F1}
\end{align}
where $L$ is the length along the $x$ direction, $f$ is a dimensionless function, $T$ is the temperature, $n=n_{\uparrow}+n_{\downarrow}=k_{F}/\pi$ is the Fermi particle number density, and $k_{F}$ is the Fermi wave vector.

Equation~(\ref{eq:F1}) implies the scaling behavior of the Helmholtz free-energy density as follows:
\begin{align}
&~~\tilde{\lambda}^{3} {\cal F}(T, n_{\uparrow}, n_{\downarrow}, a_{1D}^{}, k_{0}, \Omega)  \nonumber\\
&={\cal F}\left(\tilde{\lambda}^{2}T, \tilde{\lambda}n_{\uparrow}, \tilde{\lambda}n_{\downarrow}, \tilde{\lambda}^{-1}a_{1D}^{}, \tilde{\lambda} k_{0}, \tilde{\lambda}^{2}\Omega \right), \label{eq:F2}
\end{align} where $\tilde{\lambda}$ is a dimensionless and arbitrary parameter.

Taking the derivative of Eq.~(\ref{eq:F2}) with respect to $\tilde{\lambda}$ at $\tilde{\lambda}=1$, we have
\begin{align}
3{\cal F}&=
\left(2T \frac{\partial}{\partial T} + n_{\uparrow}\frac{\partial}{\partial n_{\uparrow}} + n_{\downarrow}\frac{\partial}{\partial n_{\downarrow}} - a_{1D}^{}\frac{\partial}{\partial a_{1D}^{}} \right. \nonumber\\
&\left.~~ + k_{0}\frac{\partial}{\partial k_{0}} + 2\Omega\frac{\partial}{\partial \Omega}\right){\cal F}. \label{eq:F3}
\end{align}

Replacing the free-energy density ${\cal F}$ on the left side of Eq.~(\ref{eq:F3}) by $n_{\uparrow}\mu_{\uparrow} + n_{\downarrow}\mu_{\downarrow} - {\cal P}$ and substituting $S=-\partial F/\partial T$ and $\mu_{\sigma}=\partial F/\partial n_{\sigma}$ into Eq.~(\ref{eq:F3}), one gets
\begin{align}
3(n_{\uparrow}\mu_{\uparrow} + n_{\downarrow}\mu_{\downarrow} - {\cal P})&=
-2T S + n_{\uparrow}\mu_{\uparrow} + n_{\downarrow}\mu_{\downarrow} \nonumber\\
&~~ - a_{1D}^{}\frac{\partial{\cal F}}{\partial a_{1D}^{}} + k_{0}\frac{\partial{\cal F}}{\partial k_{0}} + 2\Omega\frac{\partial{\cal F}}{\partial \Omega}, \label{eq:F4}
\end{align} where $\cal P$ is the pressure density, $S$ is the entropy, and $\mu_{\sigma}$ is the chemical potential with spin $\sigma$.

Using the adiabatic relations (\ref{eq:caE}), (\ref{eq:ck0}) and (\ref{eq:cOmega}), we can get the pressure relation as
\begin{eqnarray}
{\cal P} = 2{\cal E} + \frac{a_{1D}^{}C_{a}}{2mL} - \frac{k_{0}C_{\lambda}}{L} - \frac{2\Omega C_{\Omega}}{L}, \label{eq:pressure}
\end{eqnarray} where ${\cal E}=E/L$ is the energy density and we use $E=F+TS$.

\subsection{Virial theorem}\label{5.3}

For an atomic gas in a harmonic potential $V_T=m\omega^2 x^2/2$ with the trapping frequency $\omega$, the free energy can be expressed in terms of~\cite{Braaten20082,Barth2011,Cui20161}
\begin{align}
&~~F( T, \omega, a_{1D}^{}, k_{0}, \Omega, N_{\uparrow}, N_{\downarrow} ) \nonumber\\
&= \omega \tilde{f}( T/\omega, \omega/\omega, a_{1D}^{}/a_{\text{ho}}, k_{0}a_{\text{ho}}, \Omega/\omega, N_{\uparrow}, N_{\downarrow} ), \label{eq:FV1}
\end{align}
where $N=N_{\uparrow} + N_{\downarrow}$ is the particle number, $a_{\text{ho}} = \sqrt{2/(m\omega)}$ is the harmonic-oscillator length and the dimensionless function $\tilde{f}$ is dependent on the dimensionless ratios $T/\omega$, $a_{1D}^{}/a_{\text{ho}}$, $k_{0}a_{\text{ho}}$, $\Omega/\omega$, and particle numbers $N_{\uparrow}$ and $N_{\downarrow}$.

With Eq.~(\ref{eq:FV1}), we can get the scaling law
\begin{align}
&~~\tilde{\lambda} F( T, \omega, a_{1D}^{}, k_{0}, \Omega, N_{\uparrow}, N_{\downarrow} ) \nonumber\\
& = F( \tilde{\lambda}T, \tilde{\lambda}\omega, \tilde{\lambda}^{-1/2}a_{1D}^{}, \tilde{\lambda}^{1/2}k_{0}, \tilde{\lambda}\Omega, N_{\uparrow}, N_{\downarrow} ), \label{eq:FV2}
\end{align}
where $\tilde{\lambda}$ is a dimensionless and arbitrary parameter.

The derivative of Eq.~(\ref{eq:FV2}) with respect to $\tilde{\lambda}$ at $\tilde{\lambda}=1$ gives 
\begin{align}
F=
\left(T \frac{\partial}{\partial T} + \omega \frac{\partial}{\partial \omega} - \frac{1}{2}a_{1D}^{} \frac{\partial}{\partial a_{1D}^{}}+ \frac{1}{2}k_{0}\frac{\partial}{\partial k_{0}} + \Omega\frac{\partial}{\partial \Omega} \right)F. \label{eq:FV3}
\end{align}

Substituting $E=F+TS$ and $S=-\partial F/\partial T$ into Eq.~(\ref{eq:FV3}), one gets
\begin{align}
E = \left(\omega \frac{\partial}{\partial \omega} - \frac{1}{2}a_{1D}^{} \frac{\partial}{\partial a_{1D}^{}} + \frac{1}{2}k_{0}\frac{\partial}{\partial k_{0}} + \Omega\frac{\partial}{\partial \Omega} \right)E,
\end{align}
which, together with the Hellmann-Feynman theorem and the adiabatic relations (\ref{eq:caE}), (\ref{eq:ck0}) and (\ref{eq:cOmega}), gives
\begin{equation}
E = 2 \langle V_T\rangle - \frac{a_{1D}^{}C_{a}}{4m } + \frac{k_{0}C_{\lambda}}{2} + \Omega C_{\Omega}.  \label{eq:virial}
\end{equation}

\section{Large-momentum tail}\label{6}

%%%%%%%%%%%%%%%%%%%%%%%%%%%%%%%%%%%%%%%%%%%%%%%%%%
\begin{figure}
\includegraphics[width=6cm]{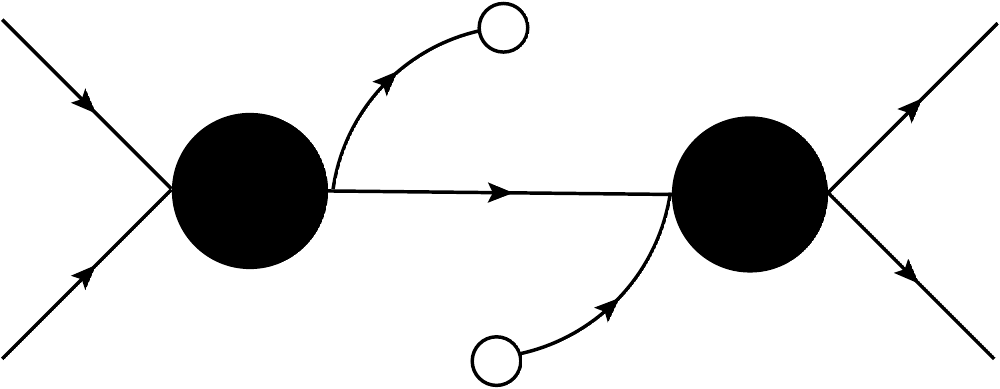}
\caption{Feynman diagram for the matrix elements of the
operator $\psi_{\sigma}^{\dag}(R+\frac{x}{2}) \psi_{\sigma'}(R-\frac{x}{2})$.
\label{fig:Distributiond}}
\end{figure}
%%%%%%%%%%%%%%%%%%%%%%%%%%%%%%%%%%%%%%%%%%%%%%%%%%

In this section, we derive the tail of the momentum distribution for 1D fermions with SOC near a broad Feshbach resonance using the OPE method~\cite{Braaten20081,Braaten20082,Cui20161,Cui20162}.

OPE is an ideal tool to explore the short-range physics.
One can expand the product of two operators as
\begin{equation}\label{ope}
\psi_{\sigma}^{\dag}(R+\frac{x}{2}) \psi_{\sigma'}(R-\frac{x}{2}) = \sum_n C_n(x) {\cal O}_n(R),
\end{equation}
where ${\cal O}_n(R)$ are the local operators and $C_n(x)$ are the short-distance coefficients.
$C_n(x)$ can be determined by calculating the matrix elements of the operators on both sides of Eq.~(\ref{ope}) in the two-body state $|Q/2 + k,\sigma; Q/2 - k,\sigma'\rangle$.

By using the Fourier transformation on both sides of Eq.~(\ref{ope}), we have the expression of momentum distribution as~\cite{Braaten20082}
\begin{align}\label{eq:rho}
n_{\sigma'\sigma}(q) = 
\int \frac{dR}{L} \int dx~e^{-iqx} \left\langle \psi_{\sigma}^{\dag}(R+\frac{x}{2}) \psi_{\sigma'}(R-\frac{x}{2}) \right\rangle,
\end{align} where $q$ is the relative momentum.

There are four types of diagrams which can be used to denote the operators on the left-hand side of the OPE equation~(\ref{ope}).
However, the only nonanalyticity comes from the diagram as shown in Fig.~\ref{fig:Distributiond}.
Therefore, one can evaluate the diagram in Fig.~\ref{fig:Distributiond} as (the derivations are given in the Appendix)
\begin{widetext}
\begin{eqnarray}\label{eq:distribution-r}
&&\left(\begin{array}{cc}
\left\langle O_s\left| \psi_{\uparrow}^{\dagger}(R+\frac{x}{2}) \psi_{\uparrow}(R-\frac{x}{2}) \right|I_s\right\rangle_{d}  & \left\langle O_s\left| \psi_{\downarrow}^{\dagger}(R+\frac{x}{2}) \psi_{\uparrow}(R-\frac{x}{2}) \right|I_s\right\rangle_{d} \\
\left\langle O_s\left| \psi_{\uparrow}^{\dagger}(R+\frac{x}{2}) \psi_{\downarrow}(R-\frac{x}{2}) \right|I_s\right\rangle_{d}  &  \left\langle O_s\left|\psi_{\downarrow}^{\dagger}(R+\frac{x}{2}) \psi_{\downarrow}(R-\frac{x}{2}) \right|I_s\right\rangle_{d} 
\end{array}\right) \nonumber\\
&=& [-iT_{s}(q_{0},Q)]^{2}\int \frac{dp dp_0}{(2\pi)^2} G(p_{0},p) \epsilon G^{T}(q_{0}-p_{0},Q-p)\epsilon^{\dagger} G(p_{0},p)  e^{-ipx}, 
\end{eqnarray}
\end{widetext} where $q_{0}^{}=Q^{2}/(4m)+k^{2}/m$ is the total energy, $Q$ is the center-of-mass momentum, and we use the Feynman rules
for the operator vertices~\cite{Braaten20082}, 
\begin{eqnarray}
\Psi^{\dag}(R+\frac{x}{2}) \Psi(R-\frac{x}{2})
&\sim& \exp\left[-ip\left(R+\frac{x}{2}\right) \right] \nonumber\\
&&\times\exp\left[ip'\left(R-\frac{x}{2}\right) \right], 
\end{eqnarray} the incoming and outgoing momenta are $p$ and $p'$ with $p=p'$.

With the Fourier transforms, we get the momentum distribution matrix as
\begin{eqnarray}\label{eq:distribution-q}
n(q)
&=& [-iT_{s}(q_{0},Q)]^{2}\int_{-\infty}^{\infty} \frac{dp_0}{2\pi} \nonumber\\
&&\times G(p_{0},q) \epsilon G^{T}(q_{0}-p_{0},Q-q)\epsilon^{\dagger} G(p_{0},q)  \nonumber\\
&=&
\left(\begin{array}{cc}
n_{\uparrow\uparrow}(q)  & n_{\uparrow\downarrow}(q) \\
n_{\downarrow\uparrow}(q)  &  n_{\downarrow\downarrow}(q)
\end{array}\right),
\end{eqnarray} where the analytical expressions for the elements of the matrix are shown in the Appendix.

Matching Eq.~(\ref{eq:distribution-q}) with Eq.~(\ref{eq:TwoBodyOperatorsR2}), we get the momentum distribution matrix in the large-$q$ limit up to the $q^{-8}$ order, 
\begin{widetext}\begin{align}
&n(q) = \frac{C_{a}}{q^{4}L} + \frac{2{\bf \hat{q}}\cdot{\bf C}_{Q1}-4k_{0}C_{a}\sigma_{z}}{q^{5}L} + \frac{2C_{R1}+10k_{0}^{2}C_{a} - 10k_{0}{\bf \hat{q}}\cdot{\bf C}_{Q1}\sigma_{z} + 5C_{Q2}/2}{q^{6}L} \nonumber\\
&+ \frac{-12k_{0}C_{R1}\sigma_{z} - 20k_{0}^{3}C_{a}\sigma_{z} + 6{\bf \hat{q}}\cdot{\bf C}_{11} + 30k_{0}^{2}{\bf \hat{q}}\cdot{\bf C}_{Q1} - 15k_{0}C_{Q2}\sigma_{z} + 5{\bf \hat{q}}\cdot{\bf C}_{Q3}/2}{q^{7}L} \nonumber\\
&+ \frac{3C_{R2} + 42k_{0}^{2}C_{R1} + 35k_{0}^{4}C_{a} - 42k_{0}{\bf \hat{q}}\cdot{\bf C}_{11}\sigma_{z} - 70k_{0}^{3}{\bf \hat{q}}\cdot{\bf C}_{Q1}\sigma_{z} + 21C_{12}/2 + 105k_{0}^{2}C_{Q2}/2 - 35k_{0}{\bf \hat{q}}\cdot{\bf C}_{Q3}\sigma_{z}/2 + 35C_{Q4}/16}{q^{8}L},
\label{eq:tail-matrix}
\end{align}\end{widetext} where we take the SOC parameters as perturbations since the strength of the SOC should be much smaller than the corresponding strength scale of the interatomic interactions, 
${\bf \hat{q}}$ is the unit vector, and the new contacts are defined as
\begin{align}
&C_{Rj} = m^{2+j} g_{1D}^{2} \int dR \nonumber\\
&~~\times\langle \psi^{\dag}_{\uparrow}(R) \psi^{\dag}_{\downarrow}(R) \left(i\partial_{t}+\frac{\partial_{R}^{2}}{4m} \right)^{j} \psi_{\downarrow}(R) \psi_{\uparrow}(R) \rangle, \\
&C_{Qn} = m^{2} g_{1D}^{2} \int dR  \nonumber\\
&~~\times\langle \psi^{\dag}_{\uparrow}(R) \psi^{\dag}_{\downarrow}(R)(-i\partial_{R})^{n} \psi_{\downarrow}(R) \psi_{\uparrow}(R) \rangle, \\
&C_{jn} = m^{2+j} g_{1D}^{2} \int dR \nonumber\\
&\times\langle \psi^{\dag}_{\uparrow}(R) \psi^{\dag}_{\downarrow}(R) \left(i\partial_{t}+\frac{\partial_{R}^{2}}{4m} \right)^{j}(-i\partial_{R})^{n}\psi_{\downarrow}(R) \psi_{\uparrow}(R) \rangle,
\end{align} $j,n=1,2,3,\cdot\cdot\cdot$.
Note that, if $n$ is an odd number, the corresponding contact is a vector.

Different from $C_{\lambda}$ and $C_{\Omega}$, $C_{Rj}$ refers to the effective range (as well as higher-order terms in the cotangent of the phase-shift expansion) which we do not discuss in our single-channel model Hamiltonian (\ref{eq:Hamiltonian-r}), $C_{Qn}$ refers to the center-of-mass momentum, and $C_{jn}$ refers to both of the effective range (as well as higher-order terms in the cotangent of the phase-shift expansion) and center-of-mass momentum.

In the laboratory frame, the single-particle Hamiltonian is given by~\cite{Zhai2012,Zhai2015}
\begin{align}\label{eq:Hamiltonian-lab}
{\cal H}_{\text{lab},0} = 
\left(\begin{array}{cc}
\frac{k^{2}}{2m}  & \Omega e^{i2k_{0}x} \\
\Omega e^{-i2k_{0}x}   &  \frac{k^{2}}{2m}
\end{array}\right).
\end{align}
By transforming $n_{\uparrow\uparrow}(q)$ to $n_{\uparrow\uparrow}(q-k_{0})$ and $n_{\downarrow\downarrow}(q)$ to $n_{\downarrow\downarrow}(q+k_{0})$, the momentum distribution can go back to the laboratory frame as discussed in Ref.~\cite{ZhangSOC2018}. Furthermore, the diagonal elements of the momentum distribution matrix in the laboratory frame are calculated as
\begin{align}\label{eq:lab-tail}
&n_{\text{lab}}(q) = \frac{C_{a}}{q^{4}L} + \frac{2{\bf \hat{q}}\cdot{\bf C}_{Q1}}{q^{5}L} + \frac{4C_{R1}+5C_{Q2}}{2q^{6}L}  \nonumber\\
&~~+ \frac{12{\bf \hat{q}}\cdot{\bf C}_{11}+5{\bf \hat{q}}\cdot{\bf C}_{Q3}}{2q^{7}L} + \frac{48C_{R2}+168C_{12}+35C_{Q4}}{16q^{8}L}.  \nonumber\\
\end{align} 
It shows an anisotropic behavior at the $q^{-5}$ and $q^{-7}$ tails due to the center-of-mass momentum.

Meanwhile, the spin-dependent momentum distribution matrix in the laboratory frame can be written as
\begin{eqnarray}\label{eq:n-lab-spin}
n_{\text{lab},\text{spin-dep}}(q)
&=& \left( -\frac{16k_{0}^{2}m\Omega C_{a}}{q^{8}L} - \frac{56k_{0}^{2}m\Omega{\bf \hat{q}}\cdot{\bf C}_{Q1}}{q^{9}L} \right)\sigma_{x} \nonumber\\
&& -\frac{16k_{0}m^{2}\Omega^{2}C_{a}}{q^{9}L} \sigma_{z}.
\end{eqnarray}
The diagonal elements of the momentum distribution matrix provide the expectation value of the atomic
number with either spin $\uparrow$ or spin $\downarrow$ with a certain relative momentum $q$. 
The off-diagonal elements indicate the mixing of
different spins.
From the above equation (\ref{eq:n-lab-spin}), it is found that the $q^{-8}$ tail in the off-diagonal terms of the momentum distribution matrix is dependent on the SOC parameters in the laboratory frame. This tail can be observed through time-of-flight measurement as a direct manifestation of the SOC effects on the many-body level.

\section{Radio-frequency spectroscopy}\label{7}

The radio-frequency spectroscopy can be used as an important experimental tool in cold atom systems and  
it can be derived from the Raman spectroscopy with zero relative
momentum between two Raman lasers. 
In order to get the radio-frequency spectroscopy, we should derive the Raman spectroscopy first as follows. 
When the transfer momentum and frequency is larger compared to the many-body scale, the Raman spectroscopy can be related to the contacts. 

We apply a Raman coupling with frequency $\omega$ and momentum $k$ which is applied to transfers fermions from the internal spin state $|\sigma\rangle$ ($\sigma=\uparrow,\downarrow$) into a third spin state $|3\rangle$. The Hamiltonian reads
\begin{align}
H_c=\sum_\sigma\Omega_{\sigma}\int dx ~e^{i(kx-\omega t)}{\cal O}_{\sigma3}(x,t)+\text{H.c.},
\end{align} where ${\cal O}_{\sigma3}(x,t) \equiv \psi_{3}^{\dag}(x,t)\psi_{\sigma}(x,t)$, $\Omega_{\sigma}$ is the radio-frequency Rabi frequency determined by the strength of the radio-frequency signal and we assume $\omega>0$.

The transition rate function $R(\omega,k)$ is given by the Fermi golden rule, which is related to the imaginary part of the time-ordered two-point correlation function~\cite{BraatenRF2010,HofmannRF2011},
\begin{align}\label{eq:ra}
\Gamma^{R}_{\sigma\sigma'}(\omega,k) &= \frac{1}{\pi} \text{Im}~\int dR \int dt~e^{i\omega t} \int dx~e^{-ikx} \nonumber\\
&\times i\left\langle {\cal T}{\cal O}_{\sigma3}(R+x,t){\cal O}_{\sigma'3}^{\dag}(R,0) \right\rangle.
\end{align} Explicitly, we have the transition rate function $R(\omega,k)$, 
\begin{align}
R(\omega,k)=2\pi\sum_{\sigma\sigma'}\Omega_\sigma\Omega_{\sigma'}^*\Gamma^{R}_{\sigma\sigma'}(\omega,k).
\end{align}

%%%%%%%%%%%%%%%%%%%%%%%%%%%%%%%%%%%%%%%%%%%%%%%%%%
\begin{figure}
\includegraphics[width=5.5cm]{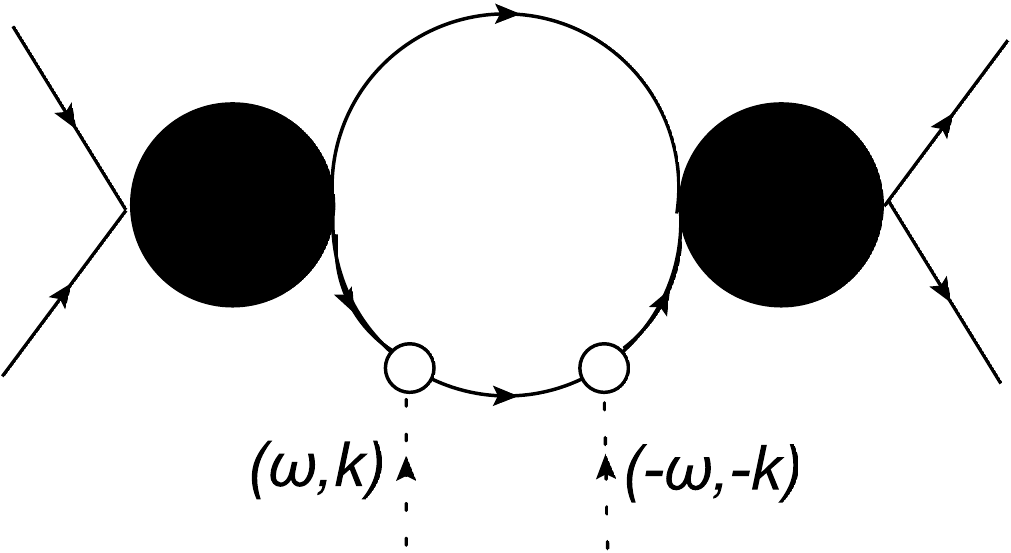}
\caption{Feynman diagram for the matrix element of $\int dt~e^{i\omega t}\int dx~e^{-ikx}{\cal T}{\cal O}_{\sigma3}(R+x,t){\cal O}_{\sigma'3}^{\dagger}(R,0)$ ($\sigma=\uparrow,\downarrow$). \label{fig:rf}}
\end{figure}
%%%%%%%%%%%%%%%%%%%%%%%%%%%%%%%%%%%%%%%%%%%%%%%%%%

Furthermore, we can evaluate the diagram in Figs.~\ref{fig:rf} as
\begin{align}\label{eq:ra-SOC}
&\Gamma^{R}_{\sigma\sigma'}(\omega,k) = \frac{1}{\pi}\text{Im}~i[-iT_{s}(E_{0},Q)]^{2} \nonumber \\
&\times \int \frac{dp dp_{0}}{(2\pi)^{2}} G_0(p_0+\omega,p+k) \nonumber \\
&\times \left[G(p_{0},p)\epsilon^{\dagger} G^{T}(E_{0}-p_{0},Q-p)\epsilon G(p_{0},p)\right]_{\sigma\sigma'},
\end{align} where $E_{0}=Q^{2}/(4m)+q^{2}/m$ is the total energy, $Q$ is the total momentum, and we have defined $G_0(p_{0}^{},p)=i/[p_{0}^{}-p^2/(2m)+i0^+]$.

Matching Eq.~(\ref{eq:ra}) with Eq.~(\ref{eq:TwoBodyOperatorsR2}) and taking the limit of $k=0$, we have the radio-frequency spectral $\Gamma^{rf}_{\sigma\sigma'}(\omega)=\Gamma^{R}_{\sigma\sigma'}(\omega,0)$ in the high-frequency limit,
\begin{align}\label{eq:rf-SOC}
\Gamma^{\text{rf}}(\omega) &= \frac{m}{2\pi}\left[ \frac{C_{a}}{(m\omega)^{5/2}} + \frac{25k_{0}^{2}C_{a} - 4C_{R1} - 10k_{0}{\bf \hat{q}}\cdot{\bf C}_{Q1}\sigma_{z} }{8(m\omega)^{7/2}} \right] \nonumber\\
&~~ - \frac{5m^{2}\Omega C_{a}}{4\pi(m\omega)^{7/2}}\sigma_{x}.
\end{align}
As shown in Eq.~(\ref{eq:rf-SOC}), it is found that the high-frequency radio-frequency spectrum contains $C_{a}$, $C_{R1}$, and ${\bf C}_{Q1}$ as expected. Especially, the presence of ${\bf C}_{Q1}$ in the radio-frequency spectrum is due to the SOC effects.

\section{Summary}\label{8}

We study the 1D two-component
Fermi gases with SOC using the single-channel model and discuss
some universal behaviors near a broad Feshbach resonance. 
Through the variation of energy with respect to the SOC parameters, two new physical quantities
are defined, the pressure relation, and the viral theorem are
derived in terms of the new SOC physical quantities. Utilizing the technique of OPE, the
tail of the momentum distribution matrix is obtained. 
The momentum distribution matrix shows a spin-dependent behavior due to the SOC, and it shows an anisotropic behavior at the $q^{-5}$ and $q^{-7}$ tails due to the center-of-mass momentum.
In the laboratory frame, the $q^{-8}$ tail in the momentum distribution can be observed through time-of-flight measurement as a direct manifestation of the SOC effects on the many-body level.
In addition, the presence of the contact related to the center-of-mass momentum in the radio-frequency spectral is also due to the SOC effects.

\section*{Acknowledgements}

We acknowledge helpful discussions with Wei Yi, Xiaoling Cui, and Lianyi He.
This work was supported by the National Natural Science Foundation of China (Grant No. 11404106).
F.Q. acknowledges support from the project funded by the China Postdoctoral Science Foundation (Grant No. 2019M662150) and SUSTech Presidential Postdoctoral Fellowship.

\appendix
\section{Appendix}\label{9}

\subsection{Derivations of Eq.~(\ref{eq:bubble})}\label{9.1}

%%%%%%%%%%%%%%%%%%%%%%%%%%%%%%%%%%%%%%%%%%%%%%%%%%
\begin{figure}
\includegraphics[width=5cm]{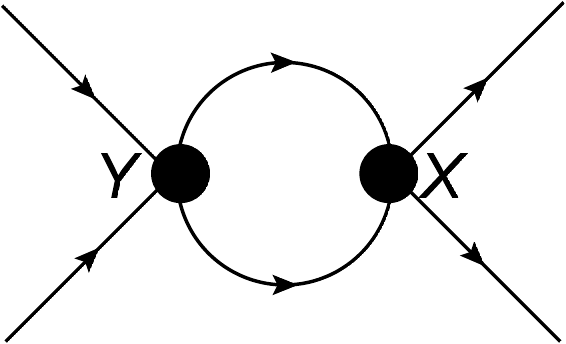}
\caption{Feynman diagram for the polarization bubble $\Pi_{s}$. Here, $X=(x_{1},t_{1})$ and $Y=(x_{2},t_{2})$. \label{fig:Bubble}}
\end{figure}
%%%%%%%%%%%%%%%%%%%%%%%%%%%%%%%%%%%%%%%%%%%%%%%%%%

The polarization bubble can be calculated by the diagram in Fig.~\ref{fig:Bubble} as
\begin{widetext}\begin{align}
\Pi_{s}(q_{0},Q) 
&= \int dXdY~\text{Tr}\left\langle {\cal T} \left[\frac{1}{2}\Psi_{f}^{T}(X)\epsilon\Psi_{f}(X)\right]\left[\frac{1}{2} \Psi_{f}^{\dagger}(Y)\epsilon^{\dagger}\Psi_{f}^{\dagger T}(Y)\right] \right\rangle \nonumber\\
&= \frac{1}{4}\int dXdY~\text{Tr}\left[ \langle {\cal T} \contraction[1.2ex]{}{A}{A\hspace{11.4ex}}{A}\Psi_{f}^{T}(X)\epsilon\contraction[2ex]{}{A}{A\hspace{11.8ex}}{A}\Psi_{f}(X) \Psi_{f}^{\dagger}(Y)\epsilon^{\dagger}\Psi_{f}^{\dagger T}(Y) \rangle + \langle {\cal T}\contraction[1.2ex]{}{A}{A\hspace{20ex}}{A}\Psi_{f}^{T}(X)\epsilon\contraction[2ex]{}{A}{A\hspace{3.2ex}}{A}\Psi_{f}(X) \Psi_{f}^{\dagger}(Y)\epsilon^{\dagger}\Psi_{f}^{\dagger T}(Y) \rangle \right] \nonumber\\
&= \frac{2}{4} \int dXdY~\text{Tr}\langle {\cal T}\contraction[1.2ex]{}{A}{A\hspace{20ex}}{A}\Psi_{f}^{T}(X)\epsilon\contraction[2ex]{}{A}{A\hspace{3.2ex}}{A}\Psi_{f}(X) \Psi_{f}^{\dagger}(Y)\epsilon^{\dagger}\Psi_{f}^{\dagger T}(Y) \rangle~(two~equivalent~contractions), \nonumber\\
&= \frac{1}{2} \int dXdY~\text{Tr}\left\langle {\cal T} \left[\Psi_{f}^{T}(X)\right]_{1a}\left(\epsilon\right)_{ab}\left[\Psi_{f}(X)\right]_{b1} \left[\Psi_{f}^{\dagger}(Y)\right]_{1m}\left(\epsilon^{\dagger}\right)_{mn}\left[\Psi_{f}^{\dagger T}(Y)\right]_{n1} \right\rangle \nonumber\\
&= \frac{1}{2} \int dXdY~\text{Tr}\left\langle {\cal T} \left[\Psi_{f}^{T}(X)\right]_{1a}\left[\Psi_{f}^{\dagger T}(Y)\right]_{n1}\left(\epsilon\right)_{ab}\left[\Psi_{f}(X)\right]_{b1} \left[\Psi_{f}^{\dagger}(Y)\right]_{1m}\left(\epsilon^{\dagger}\right)_{mn} \right\rangle \nonumber\\
&= \frac{1}{2} \int dXdY~\text{Tr}\left\langle {\cal T} \left[\Psi_{f}(X)\right]_{a1}\left[\Psi_{f}^{\dagger}(Y)\right]_{1n} \right\rangle\left(\epsilon\right)_{ab}\left[G(X-Y)\right]_{bm}\left(\epsilon^{\dagger}\right)_{mn} \nonumber\\
&= \frac{1}{2} \int dXdY~\text{Tr}\left[G(X-Y)\right]_{an}\left(\epsilon\right)_{ab}\left[G(X-Y)\right]_{bm}\left(\epsilon^{\dagger}\right)_{mn} \nonumber\\
&= \frac{1}{2} \int dXdY~\text{Tr}\left[G^{T}(X-Y)\right]_{na}\left(\epsilon\right)_{ab}\left[G(X-Y)\right]_{bm}\left(\epsilon^{\dagger}\right)_{mn} \nonumber\\
&= \int dXdY~\frac{1}{2}\text{Tr}\left[ G^{T}(X-Y)\epsilon G(X-Y)\epsilon^{\dagger} \right],
\end{align}\end{widetext} where $\int dXdY = \frac{1}{2} \int d(X+Y)d(X-Y)$, we label the field operator $\Psi=\Psi_{s}+\Psi_{f}$ for outline as $\Psi_{s}$ and inner lines as $\Psi_{f}$, $\Psi_{f}=(\psi_{\uparrow},\psi_{\downarrow})^{T}$, $X=(x_{1},t_{1})$, $Y=(x_{2},t_{2})$, $q_{0}=Q^{2}/(4m)+p^{2}/m$, and we use the definition $G(X-Y)\equiv\langle {\cal T} \Psi_{f}(X)\Psi_{f}^{\dagger}(Y)\rangle$.

\subsection{Derivations of Eq.~(\ref{eq:g1D})}\label{9.2}

In the absence of the Raman coupling, the polarization bubble is given by
\begin{align}
\Pi_{s}(q_{0},Q) &= \int\frac{dpdp_{0}}{(2\pi)^2} \frac{i}{p_{0}-(Q/2+p)^{2}/(2m)+i0^{+}}\nonumber\\
&~~\times\frac{i}{q_{0}-p_{0}-(Q/2-p)^{2}/(2m)+i0^{+}} \nonumber\\
&= \frac{m}{2k}, 
\end{align} where $q_{0}=Q^{2}/(4m) + k^{2}/m$.

The even-wave scattering amplitude can be written as~\cite{Barth2011,Olshanii1998} $f_{1D}(k) = -1/(1+i\cot\delta_{k}) \simeq -1/(1+ika_{1D})$, where $\delta_{k}$ is the scattering phase shift.
With $T_{s}(k)=ikf_{1D}(k)/m_{r}$, one can derive Eq.~(\ref{eq:g1D}) in the main text.

\subsection{Derivations of Eq.~(\ref{eq:distribution-r})}\label{9.3}

%%%%%%%%%%%%%%%%%%%%%%%%%%%%%%%%%%%%%%%%%%%%%%%%%%
\begin{figure}
\includegraphics[width=8cm]{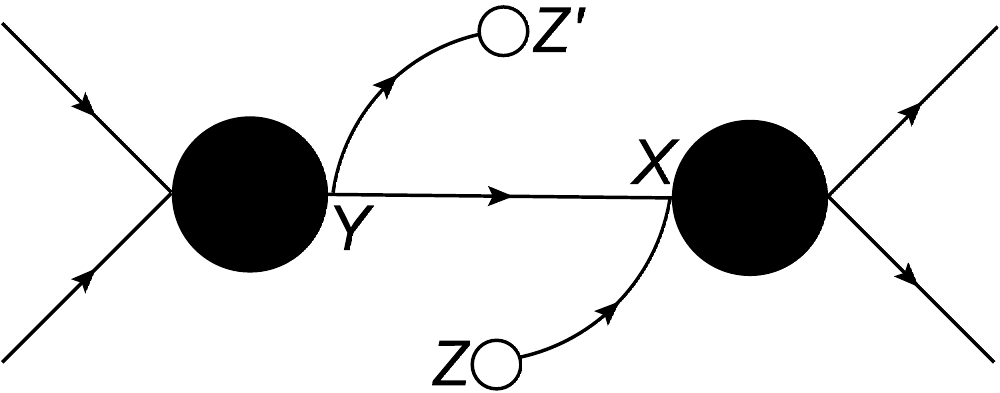}
\caption{Feynman diagram for the matrix of the
operator $\Psi^{\dagger T}(Z)\Psi^{T}(Z')$. Here, $X=(x_{1},t_{1})$, $Y=(x_{2},t_{2})$, $Z=(x_{3},t_{3})$, and $Z'=(x_{4},t_{4})$. \label{fig:DistributionAppendix}}
\end{figure}
%%%%%%%%%%%%%%%%%%%%%%%%%%%%%%%%%%%%%%%%%%%%%%%%%%

As shown in Fig.~\ref{fig:DistributionAppendix}, the vacuum expectation of the operator $\Psi^{\dagger T}(Z)\Psi^{T}(Z')$ is calculated as follows: 
\begin{widetext}
\begin{eqnarray}
&&\left\langle O_s\left| \Psi^{\dagger T}(Z)\Psi^{T}(Z') \right|I_s\right\rangle_{d} = \left(\begin{array}{cc}
\left\langle O_s\left| \psi_{\uparrow}^{\dagger}(Z) \psi_{\uparrow}(Z') \right|I_s\right\rangle_{d}  & \left\langle O_s\left| \psi_{\uparrow}^{\dagger}(Z) \psi_{\downarrow}(Z') \right|I_s\right\rangle_{d} \\
\left\langle O_s\left| \psi_{\downarrow}^{\dagger}(Z) \psi_{\uparrow}(Z') \right|I_s\right\rangle_{d}  &  \left\langle O_s\left|\psi_{\downarrow}^{\dagger}(Z) \psi_{\downarrow}(Z') \right|I_s\right\rangle_{d}
\end{array}\right) \nonumber\\  
&=& [-iT_{s}(q_{0},Q)]^{2}\int dXdY~\left\langle {\cal T} \left[\frac{1}{2} \Psi_{f}^{T}(X)\epsilon\Psi_{f}(X) \right]\Psi_{f}^{\dagger T}(Z)\Psi_{f}^{T}(Z')   \left[\frac{1}{2}\Psi_{f}^{\dagger}(Y)\epsilon^{\dagger}\Psi_{f}^{\dagger T}(Y)\right]  \right\rangle \nonumber \\
&=& [-iT_{s}(q_{0},Q)]^{2}\int dXdY~\left\langle {\cal T}\Psi_{f}^{\dagger T}(Z) \left[\frac{1}{2} \Psi_{f}^{T}(X)\epsilon\Psi_{f}(X) \right] \left[\frac{1}{2}\Psi_{f}^{\dagger}(Y)\epsilon^{\dagger}\Psi_{f}^{\dagger T}(Y)\right]  \Psi_{f}^{T}(Z')  \right\rangle \nonumber \\
&=& \frac{[-iT_{s}(q_{0},Q)]^{2}}{4}\int dXdY \nonumber\\
&&\left[\langle {\cal T} \contraction[1.2ex]{}{A}{A\hspace{3.9ex}}{A}\Psi_{f}^{\dagger T}(Z)\Psi_{f}^{T}(X)\epsilon\contraction[1.2ex]{}{A}{A\hspace{3.3ex}}{A}\Psi_{f}(X) \Psi_{f}^{\dagger}(Y)\epsilon^{\dagger}\contraction[1.2ex]{}{A}{A\hspace{4.1ex}}{A}\Psi_{f}^{\dagger T}(Y) \Psi_{f}^{T}(Z') \rangle 
+ \langle {\cal T} \contraction[1.2ex]{}{A}{A\hspace{3.9ex}}{A}\Psi_{f}^{\dagger T}(Z)\Psi_{f}^{T}(X)\epsilon\bcontraction[1.2ex]{}{A}{A\hspace{11.9ex}}{A}\Psi_{f}(X) \contraction[1.2ex]{}{A}{A\hspace{12.7ex}}{A}\Psi_{f}^{\dagger}(Y)\epsilon^{\dagger}\Psi_{f}^{\dagger T}(Y) \Psi_{f}^{T}(Z') \rangle \right. \nonumber\\
&&\left.+ \langle {\cal T} \contraction[1.2ex]{}{A}{A\hspace{12.1ex}}{A}\Psi_{f}^{\dagger T}(Z)\bcontraction[1.2ex]{}{A}{A\hspace{11.5ex}}{A}\Psi_{f}^{T}(X)\epsilon\Psi_{f}(X) \Psi_{f}^{\dagger}(Y)\epsilon^{\dagger}\contraction[1.2ex]{}{A}{A\hspace{4ex}}{A}\Psi_{f}^{\dagger T}(Y) \Psi_{f}^{T}(Z') \rangle 
+ \langle {\cal T} \contraction[1.2ex]{}{A}{A\hspace{12.1ex}}{A}\Psi_{f}^{\dagger T}(Z)\bcontraction[1.2ex]{}{A}{A\hspace{20ex}}{A}\Psi_{f}^{T}(X)\epsilon\Psi_{f}(X) \contraction[1.2ex]{}{A}{A\hspace{12.6ex}}{A}\Psi_{f}^{\dagger}(Y)\epsilon^{\dagger}\Psi_{f}^{\dagger T}(Y) \Psi_{f}^{T}(Z') \rangle \right] \nonumber \\
&=& [-iT_{s}(q_{0},Q)]^{2}\int dXdY~\langle {\cal T} \contraction[1.2ex]{}{A}{A\hspace{3.9ex}}{A}\Psi_{f}^{\dagger T}(Z)\Psi_{f}^{T}(X)\epsilon\contraction[1.2ex]{}{A}{A\hspace{3.3ex}}{A}\Psi_{f}(X) \Psi_{f}^{\dagger}(Y)\epsilon^{\dagger}\contraction[1.2ex]{}{A}{A\hspace{4.1ex}}{A}\Psi_{f}^{\dagger T}(Y) \Psi_{f}^{T}(Z') \rangle~(four~equivalent~contractions),  \nonumber \\
&=& [-iT_{s}(q_{0},Q)]^{2}\int dXdY~\left\langle {\cal T}\left[\Psi_{f}^{\dagger T}(Z)\right]_{j1} \left[\Psi_{f}^{T}(X)\right]_{1m}\left(\epsilon\right)_{mn}\left[\Psi_{f}(X)\right]_{n1} \left[\Psi_{f}^{\dagger}(Y)\right]_{1a}\left(\epsilon^{\dagger}\right)_{ab}\left[\Psi_{f}^{\dagger T}(Y)\right]_{b1} \left[\Psi_{f}^{T}(Z')\right]_{1j}  \right\rangle  \nonumber \\
&=& [-iT_{s}(q_{0},Q)]^{2}\int dXdY~\left\langle {\cal T}\left[\Psi_{f}(X)\right]_{m1}\left[\Psi_{f}^{\dagger}(Z)\right]_{1j} \left(\epsilon\right)_{mn}\left[\Psi_{f}(X)\right]_{n1}\left[\Psi_{f}^{\dagger}(Y)\right]_{1a} \left(\epsilon^{\dagger}\right)_{ab}\left[\Psi_{f}(Z')\right]_{j1} \left[\Psi_{f}^{\dagger}(Y)\right]_{1b} \right\rangle  \nonumber \\
&=& [-iT_{s}(q_{0},Q)]^{2}\int dXdY~\left[G(X-Z)\right]_{mj}\left(\epsilon\right)_{mn}\left[G(X-Y)\right]_{na} \left(\epsilon^{\dagger}\right)_{ab}\left[G(Z'-Y)\right]_{jb} \nonumber \\
&=& [-iT_{s}(q_{0},Q)]^{2}\int dXdY~\left[G^{T}(X-Z)\right]_{jm}\left(\epsilon\right)_{mn}\left[G(X-Y)\right]_{na} \left(\epsilon^{\dagger}\right)_{ab}\left[G^{T}(Z'-Y)\right]_{bj} \nonumber \\
&=& [-iT_{s}(q_{0},Q)]^{2}\int dXdY~G^{T}(X-Z) \epsilon G(X-Y)\epsilon^{\dagger} G^{T}(Z'-Y).
\end{eqnarray}
\end{widetext} 

Therefore, we have
\begin{eqnarray}\label{eq:Psi}
&&\left(\begin{array}{cc}
\left\langle O_s\left| \psi_{\uparrow}^{\dagger}(Z) \psi_{\uparrow}(Z') \right|I_s\right\rangle_{d}  & \left\langle O_s\left| \psi_{\uparrow}^{\dagger}(Z) \psi_{\downarrow}(Z') \right|I_s\right\rangle_{d} \\
\left\langle O_s\left| \psi_{\downarrow}^{\dagger}(Z) \psi_{\uparrow}(Z') \right|I_s\right\rangle_{d}  &  \left\langle O_s\left|\psi_{\downarrow}^{\dagger}(Z) \psi_{\downarrow}(Z') \right|I_s\right\rangle_{d}
\end{array}\right)^{T} \nonumber \\
&=& \left[ \left\langle O_s\left| \Psi^{\dagger T}(Z)\Psi^{T}(Z') \right|I_s\right\rangle_{d} \right]^{T} \nonumber \\
&=& [-iT_{s}(q_{0},Q)]^{2}\nonumber \\
&&\times\int dXdY~ G(Z'-Y)\epsilon G^{T}(X-Y)\epsilon^{\dagger}G(X-Z),
\end{eqnarray} where we use $(ABCDE)^{T}=E^{T}D^{T}C^{T}B^{T}A^{T}$.

\subsection{Elements of the momentum distribution matrix Eq.~(\ref{eq:distribution-q})}\label{9.4}

\begin{widetext}
\begin{eqnarray}
n_{\uparrow\uparrow}(q) &=& -[T_{s}(q_{0},Q)]^{2}\int_{-\infty}^{\infty} \frac{dp_0}{2\pi}\left[ G_{\uparrow\uparrow}^{2}(p_{0},q)G_{\downarrow\downarrow}(q_{0}-p_{0},Q-q) + G_{\uparrow\uparrow}(q_{0}-p_{0},Q-q)G_{\uparrow\downarrow}(p_{0},q)G_{\downarrow\uparrow}(p_{0},q) \right. \nonumber\\
&&\left. -G_{\uparrow\uparrow}(p_{0},q)G_{\uparrow\downarrow}(p_{0},q)G_{\downarrow\uparrow}(q_{0}-p_{0},Q-q) - G_{\uparrow\uparrow}(p_{0},q)G_{\downarrow\uparrow}(p_{0},q)G_{\uparrow\downarrow}(q_{0}-p_{0},Q-q) \right],\\
n_{\uparrow\downarrow}(q) &=& -[T_{s}(q_{0},Q)]^{2}\int_{-\infty}^{\infty} \frac{dp_0}{2\pi}\left[ G_{\uparrow\uparrow}(p_{0},q)G_{\downarrow\downarrow}(q_{0}-p_{0},Q-q)G_{\uparrow\downarrow}(p_{0},q) + G_{\uparrow\uparrow}(q_{0}-p_{0},Q-q)G_{\downarrow\downarrow}(p_{0},q)G_{\uparrow\downarrow}(p_{0},q) \right. \nonumber\\
&&\left. -G_{\uparrow\uparrow}(p_{0},q)G_{\downarrow\downarrow}(p_{0},q)G_{\uparrow\downarrow}(q_{0}-p_{0},Q-q) - G_{\uparrow\downarrow}^{2}(p_{0},q)G_{\downarrow\uparrow}(q_{0}-p_{0},Q-q) \right],\\
n_{\downarrow\uparrow}(q) &=& -[T_{s}(q_{0},Q)]^{2}\int_{-\infty}^{\infty} \frac{dp_0}{2\pi}\left[ G_{\uparrow\uparrow}(p_{0},q)G_{\downarrow\downarrow}(q_{0}-p_{0},Q-q)G_{\downarrow\uparrow}(p_{0},q) + G_{\uparrow\uparrow}(q_{0}-p_{0},Q-q)G_{\downarrow\downarrow}(p_{0},q)G_{\downarrow\uparrow}(p_{0},q) \right. \nonumber\\
&&\left. -G_{\uparrow\uparrow}(p_{0},q)G_{\downarrow\downarrow}(p_{0},q)G_{\downarrow\uparrow}(q_{0}-p_{0},Q-q) - G_{\downarrow\uparrow}^{2}(p_{0},q)G_{\uparrow\downarrow}(q_{0}-p_{0},Q-q) \right],\\
n_{\downarrow\downarrow}(q) &=& -[T_{s}(q_{0},Q)]^{2}\int_{-\infty}^{\infty} \frac{dp_0}{2\pi}\left[ G_{\downarrow\downarrow}^{2}(p_{0},q)G_{\uparrow\uparrow}(q_{0}-p_{0},Q-q) + G_{\downarrow\downarrow}(q_{0}-p_{0},Q-q)G_{\downarrow\uparrow}(p_{0},q)G_{\uparrow\downarrow}(p_{0},q) \right. \nonumber\\
&&\left. -G_{\downarrow\downarrow}(p_{0},q)G_{\uparrow\downarrow}(p_{0},q)G_{\downarrow\uparrow}(q_{0}-p_{0},Q-q) - G_{\downarrow\downarrow}(p_{0},q)G_{\downarrow\uparrow}(p_{0},q)G_{\uparrow\downarrow}(q_{0}-p_{0},Q-q) \right].
\end{eqnarray}
\end{widetext}

\subsection{Perturbations of the elements of the single-particle propagator matrix}\label{9.5}

We use the expansions with a small $\Omega$ as a perturbation, 
\begin{align}
&G_{\uparrow\uparrow}(p_{0},q) \approx \frac{i}{ p_{0} - \frac{(q + k_{0})^{2}}{2m} + i0^{+} }\nonumber\\
&\times\left\{1 + \frac{\Omega^{2}}{ \left[p_{0} - \frac{(q + k_{0})^{2}}{2m} + i0^{+} \right] \left[p_{0} - \frac{(q - k_{0})^{2}}{2m} + i0^{+}\right] } \right\}, 
\end{align}
\begin{align}
&G_{\uparrow\downarrow}(p_{0},q) = G_{\downarrow\uparrow}(p_{0},q) \approx \nonumber\\
& \frac{i\Omega}{\left[ p_{0} - \frac{(q - k_{0})^{2}}{2m} + i0^{+} \right]\left[ p_{0} - \frac{(q + k_{0})^{2}}{2m} + i0^{+} \right] }\nonumber\\
&\times\left\{1 + \frac{\Omega^{2}}{ \left[p_{0} - \frac{(q - k_{0})^{2}}{2m} + i0^{+} \right] \left[p_{0} - \frac{(q + k_{0})^{2}}{2m} + i0^{+}\right]} \right\},
\end{align}
\begin{align}
&G_{\downarrow\downarrow}(p_{0},q) \approx \frac{i}{ p_{0} - \frac{(q - k_{0})^{2}}{2m} + i0^{+} }\nonumber\\
&\times\left\{1 + \frac{\Omega^{2}}{ \left[p_{0} - \frac{(q - k_{0})^{2}}{2m} + i0^{+} \right] \left[p_{0} - \frac{(q + k_{0})^{2}}{2m} + i0^{+}\right]} \right\}.
\end{align}

\subsection{One-dimensional Fourier transforms}\label{9.6}

The 1D Fourier transforms are~\cite{Barth2011,Fourier1D}
\begin{align}
\int dx~|x|^{\alpha}e^{-iqx} = -2\alpha!\sin\left(\frac{\pi\alpha}{2} \right)\frac{1}{|q|^{\alpha+1}},
\end{align} with $\alpha$ as an odd number, and 
\begin{align}
\int dx~\text{sgn}(x)|x|^{\beta}e^{-iqx} = -2i\beta!\cos\left(\frac{\pi\beta}{2} \right)\frac{\text{sgn}(q)}{|q|^{\beta+1}},
\end{align} with $\beta$ as an even number and $\text{sgn}(x)$ as the sign function.

%\end{CJK*}
\end{document}